# Comparison of time-resolved photoluminescence and deep-level transient spectroscopy defect evaluations in an InAs *nBn* detector subjected to *in situ* and *ex situ* 63 MeV proton irradiation

Rigo A. Carrasco ; Christopher P. Hains ; Nathan Gajowski ; Alexander T. Newell; Julie V. Logan ; Zinah M. Alsaad ; Preston T. Webster ; Christian P. Morath ; Diana Maestas ; Aaron J. Muhowski ; Samuel D. Hawkins ; Evan M. Anderson

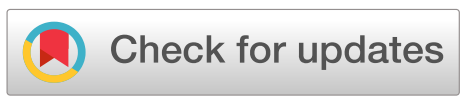

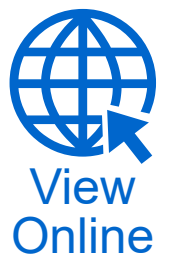

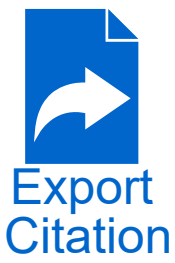

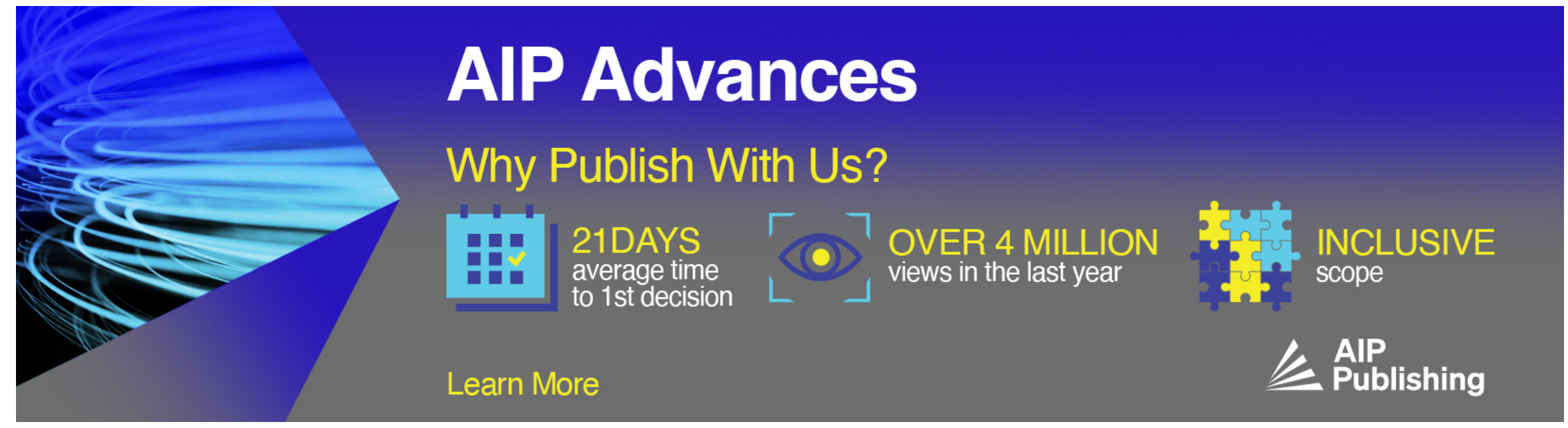

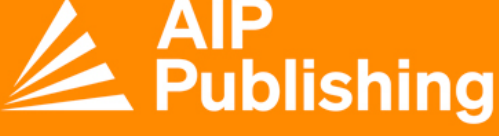

# Comparison of time-resolved photoluminescence and deep-level transient spectroscopy defect evaluations in an InAs *nBn* detector subjected to *in situ* and *ex situ* 63 MeV proton irradiation

Rigo A. Carrasco,[1,a)] Christopher P. Hains,[1,2] Nathan Gajowski,[3] Alexander T. Newell,[1] Julie V. Logan,[1] Zinah M. Alsaad,[1] Preston T. Webster,[1] Christian P. Morath,[1] Diana Maestas,[1] Aaron J. Muhowski,[4] Samuel D. Hawkins,[4] and Evan M. Anderson[4]

## AFFILIATIONS

[1]Air Force Research Laboratory, Space Vehicles Directorate, Kirtland AFB, New Mexico 87117, USA
[2]BlueHalo LLC, Albuquerque, New Mexico 87123, USA
[3]Department of Electrical Engineering, The Ohio State University, Columbus, Ohio 43210, USA
[4]Sandia National Laboratories, Albuquerque, New Mexico 87185, USA

[a)]**Author to whom correspondence should be addressed:** rigo.carrasco.1@spaceforce.mil

## ABSTRACT

Deep-level transient spectroscopy and temperature-dependent time-resolved photoluminescence experiments are performed on identical InAs *nBn* photodetector structures as a function of *in situ* and *ex situ* 63 MeV proton irradiation to assess their generation and recombination dynamics. Pre-irradiation, the *n*-type InAs absorbing region, exhibits a steadily increasing minority carrier lifetime with increasing temperature, providing evidence that excited minority carriers may be recombining via shallow defect levels. From deep-level transient spectroscopy, two features are found between 10 and 275 K: a low temperature broad "shoulder," which suggests emission from multiple shallow electron defect levels with energies <29 meV and a high temperature minimum occurring at ∼230 K with an activation energy of 539 meV, which suggests a defect in the barrier layer in the device. Two similar *nBn* detectors are then subjected to 63 MeV proton irradiation in step doses and measured between steps. One experiment is performed *in situ* with an *nBn* held at ∼10 K during dosing, and the other experiment is performed *ex situ* with a similar *nBn* held at room temperature for dosing. The *ex situ* dosing results in an evaluation of the defect introduction rate that is three to four times lower than *in situ* due to partial annealing of the proton-induced displacement damage at room temperature. The results of these two experiments are then compared with the dose-dependent recombination rate analysis, resulting in an estimated recombination defect cross section of $1.6 \times 10^{-13}$ cm$^2$ for the shallow shoulder defect.

## I. INTRODUCTION

A semiconductor material's technology readiness for imaging is largely dictated by its detection capability. This involves a holistic examination of the material's detector dark current and quantum efficiency, where the two metrics determine the material's imaging sensitivity. Some of the most significant detriments to a detector's performance potential are crystalline and impurity defects that introduce energy levels within a material's bandgap and act as recombination centers. These centers lead to a decrease in the minority carrier lifetime through Shockley–Read–Hall (SRH) recombination,[1] translating to an unwanted decrease in quantum efficiency and an increase in dark current, negatively impacting the detector's overall sensitivity. In fact, a material candidate's intrinsic carrier concentration, absorption coefficient, and limiting generation-recombination mechanisms have repeatedly been shown to determine its ultimate performance potential.[2,3] Both carrier concentration and absorption coefficient are intrinsic to a semiconductor's band structure. Therefore, reducing the recombination rate (increasing the minority carrier lifetime) by eliminating defects until it reaches one of the material's intrinsic recombination mechanisms, either Auger or radiative recombination, leaves a significant portion of device performance improvement potential.[4,5] This motivates the need to fully understand the nature of the most detrimental SRH defect recombination centers.

Defect recombination centers may either be native to growth, due to non-ideal material growth conditions, or generated extrinsically as is the case when the material is subjected to high energy irradiation, which occurs for space-based sensors.[6–9] Thus, there is interest in investigating both a detector material candidate's minority carrier lifetimes and defect character as grown as well as a function of high energy irradiation. Furthermore, given the difficulty and costs associated with performing an *in situ* irradiation experiment wherein the irradiated device is held at its cryogenic operating temperature (and bias, if applicable), it is also worthwhile to compare *in situ* to *ex situ* or "bag test" experiments, where the part is simply dosed at room temperature and characterized at its cryogenic operating temperature at a later time. Measurements such as these guide the path to new device growth and design avenues to be paved for performance improvement.

A direct-bandgap semiconductor material's minority carrier lifetime can be measured via time-resolved photoluminescence (TRPL).[8,10] Examining the temperature-dependent low-excitation minority carrier lifetime allows for a recombination rate analysis and extraction of the dominating recombination mechanisms across the temperature range of interest.[11,12] Through this method, fundamental material parameters can be extracted as expectations for the material's temperature-dependent SRH,[1] Auger,[13,14] and radiative lifetimes.[15] However, this approach requires a least-squares fit of the recombination mechanisms' functional forms to the temperature-dependent minority carrier lifetimes and is subject to correlation between fit parameters or underfitting that may still result in a seemingly good fit. For example, associating a single SRH defect may be sufficient in a recombination rate analysis for a good fit when in fact, there may be multiple recombination centers at different defect levels within the material's bandgap that appear to act as a single defect.[16] Thus, while a recombination rate analysis may provide an initial assessment of what dominates the minority carrier lifetime, the path toward performance improvements still requires a deeper understanding that cannot be provided by minority carrier lifetime measurements. More specifically, the energy levels of defects in a material system impact the detector's voltage-dependent depletion current magnitude.[9,17] Therefore, a second characterization technique is needed to connect the temperature-dependent minority carrier lifetime (extracted either from TRPL, or other lifetime measurements) to material trap energy levels and trap concentrations.

To resolve the defect character in semiconductors with higher fidelity, deep-level transient spectroscopy (DLTS)[18] is a widely accepted technique for resolving defects in semiconductors and providing feedback for improving semiconductor device performance.[19] The popularity of DLTS grew with its use over the past decades on materials such as Si (bandgap energy $E_g = 1.1$ eV) and GaAs ($E_g = 1.4$ eV), such that comprehensive libraries of native, impurity-induced and irradiation-induced defects exist.[19,20] However, for context, defect activation energies ($\Delta E_a$) in both these materials are considered shallow when $\Delta E_a \leq 413$ meV, which are roughly the size of bandgaps ($E_g \leq 413$ meV) of materials used to detect mid-wave infrared light (wavelength $\geq 3\,\mu$m).

DLTS has been demonstrated in the past on narrow bandgap semiconductors such as HgCdTe[21–24] and narrow-gap III–V barrier infrared devices.[25,26] Some correspondence has even been made with other detector metrics, such as dark current,[24] and steady state photoluminescence spectroscopy experiments.[26] In the case of III–V barrier infrared detectors, these experiments have been performed on devices with highly mismatched substrates, so it may be difficult to discern which defects are intrinsic in the absorber material, and which defects are induced from interfacial misfit dislocations.[25,26] There is also very little work on *in situ* irradiation effects on DLTS due to the naturally high costs of such an experiment. Thus, a comprehensive analysis on a device with minimal crystalline strain that ties *in situ* DLTS, a technique which examines the defect character via emission of either the minority or majority carriers depending on the injection bias condition, to a complementary quantitative experiment such as TRPL, which in contrast examines defects via carrier recombination, may ultimately provide a better connection between material-level assessments and the higher level device properties of the detector such as dark current and quantum efficiency.

In this work, both DLTS and TRPL are performed on an InAs *nBn* detector structure grown by molecular beam epitaxy. The results extracted from both measurements are used to determine the defect character in both the emission and recombination dynamics of injected carriers in the *nBn*. The device is then subjected to 63 MeV proton radiation and DLTS characterization is performed on a device *in situ* starting at 10 K, which limits the effects of thermal annealing. To demonstrate why this *in situ* experiment is necessary to understand the defect dynamics and examine the extent of thermal annealing, an equivalent InAs *nBn* device is then characterized *ex situ*, where samples are dosed at room temperature and then measured following irradiation.

## II. EXPERIMENTAL PROCEDURE

### A. Sample growth and electrical measurements

The InAs *nBn* device structure is grown by molecular beam epitaxy on an *n*-type InAs substrate in the inverted *nBn* orientation (i.e., the absorber is grown on the barrier). First, a $1 \times 10^{18}$ cm$^{-3}$ n$^{+}$ InAs bottom contact layer is grown with a Si doping profile similar to the design described in Ref. 27 on the InAs substrate. Then, a 100 nm thick unintentionally doped, lattice-matched AlAsSb barrier is grown, followed by a 4 $\mu$m thick InAs:Si absorber doped to a target of $1 \times 10^{16}$ cm$^{-3}$, while the doping of the final 150 nm is graded to $1 \times 10^{18}$ cm$^{-3}$ to create a top contact layer. A cross section schematic is illustrated in Fig. 1. InAs mesa devices are then fabricated in a cleanroom using standard photolithography processes. Device mesas are isolated by etching to the barrier in a solution of citric acid and $H_2O_2$, followed by an etch in dilute HF through the barrier to reach the bottom contact layer. Ohmic contact metals are deposited by electron beam evaporation and patterned through a lift-off process. Finally, devices are wire-bonded to a pinned grid array (PGA) chip carrier.

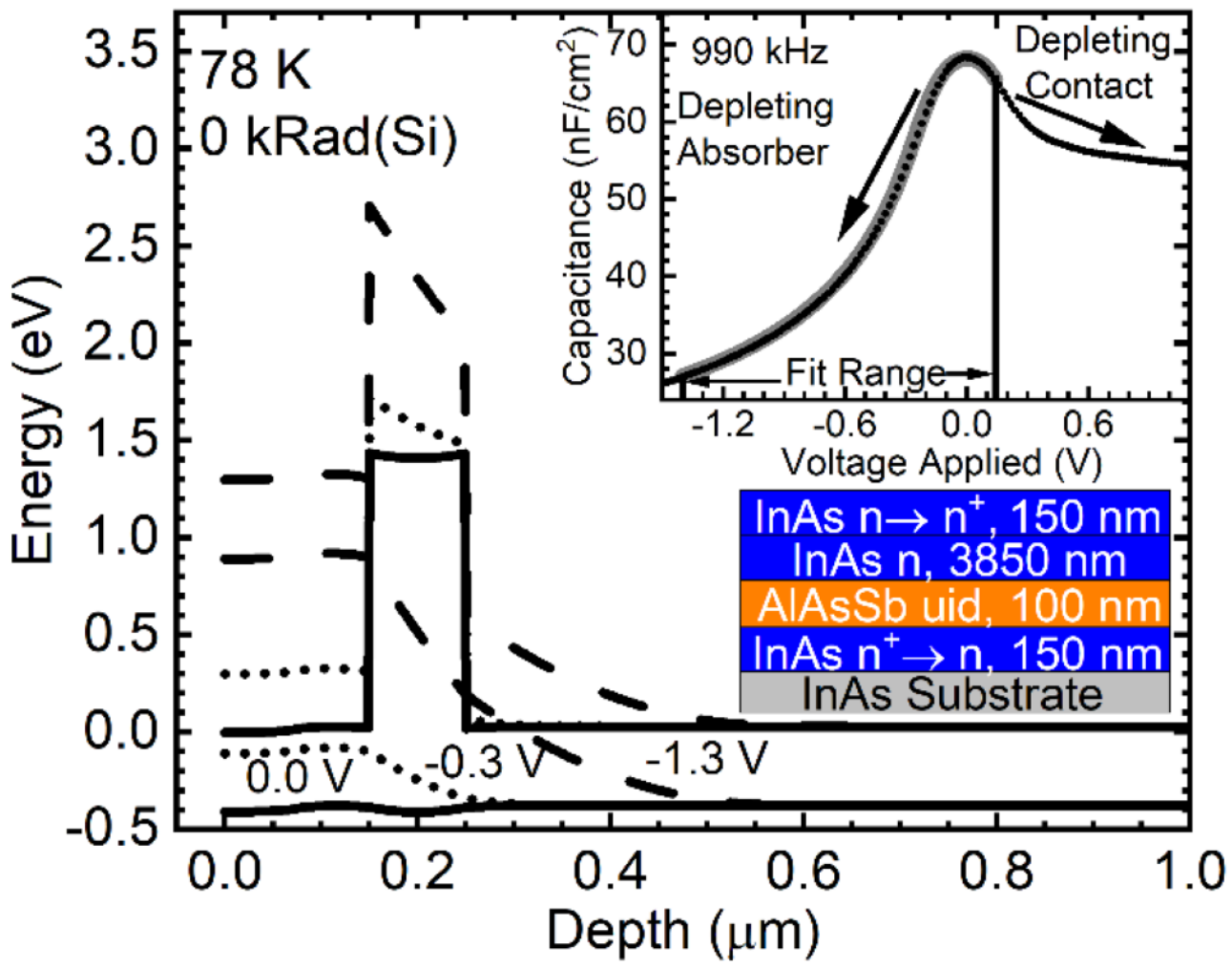

**FIG. 1.** Representative band diagrams and cross section schematic of the InAs *nBn* detector structure with a uniformly doped absorber at 0 V (solid lines), −0.3 V (dotted lines), and −1.3 V (dashed lines). The inset plot is the corresponding measured CV curve (black data points) and best fit (solid gray curve). Fit results from the inset provide inputs to the simulated band diagrams and are recorded in Table I.

The device is loaded into a pour-fill, open-cycle cryostat, where the cryogen (either $LN_2$ or LHe) is either poured in a liquid nitrogen reservoir or fed directly from a liquid helium transfer line. Thermal contact to the cryostat coldfinger is established with a thin layer of indium foil between the chip carrier and an aluminum block mounted onto the cold finger, where the temperature is controlled by a Lakeshore 336 temperature controller. Steady-state capacitance–voltage (CV) and DLTS scans are performed using a Zurich instruments medium frequency impedance analyzer (MFIA) equipped with a 5 MHz key to allow for high tunability in CV and DLTS measurements. Dark current measurements are performed with a Keithley 236 source-measure unit to determine ideal DLTS bias conditions. The MFIA allows a maximum current compliance of 10 mA, so a 500 $\mu$m mesa side-length is chosen as a reasonable compromise between having a large, measurable capacitance while also staying well below the highest possible compliance limit.

After characterization is performed for the pre-irradiation [0 kRad(Si)] condition, proton irradiation is performed at the Crocker Nuclear Laboratory at the University of California Davis.[28] The cyclotron is tuned to irradiate the detectors at 63 MeV in order to introduce a spatially uniform damage profile through the depth of the device. A correspondence of 100 kRad(Si) to $7.5 \times 10^{11}$ $H^+/cm^2$ proton fluence is determined. The detectors are held at a temperature of ∼10 K in the cryostat to prevent any annealing during doses and then capacitance transients are acquired from 10 to 130 K after each dose, acknowledging that minor thermally activated annealing may occur over the course of the DLTS acquisition. To investigate this, after the final 500 kRad dose and its associated post-rad 10–130 K DLTS sweep, a subsequent 10–130 K anneal sweep is performed. It is found that the spectra shape had changed, where values differed up to ∼10%, indicating that there is a small degree of measurement-induced annealing to be considered in the analysis in Sec. III A, which will be further investigated in a future study. Proton irradiation is delivered into the cryostat through a thin aluminum window blank, and GEANT4[29] simulations confirm that the thickness of the window has negligible effect on the non-ionizing energy loss in the InAs system. In parallel with the *in situ* step dosing experiment, a separate InAs *nBn* is exposed to step dosing at room temperature, where the transients are acquired in between room temperature dosing steps to compare the differences between *ex situ* room temperature experiments and *in situ* experiments.

### *1. Determining DLTS bias conditions*

To determine the optimal DLTS bias conditions for the InAs *nBn*, the bias-dependent energy band diagram of the *nBn* needs to be evaluated alongside bias-dependent capacitance–voltage and dark current profiles. CV measurements are performed with a 100 mV modulation amplitude and 990 kHz frequency to match the amplitude and frequency of the DLTS transients. Similar CV measurements are performed at 100 kHz and are found to have a minimal effect on the CV values over the temperatures of interest (T < 200 K). Using the model introduced in Ref. 30, absorber doping, barrier thickness, and barrier doping are extracted from the CV curve. The results are then used as inputs to simulate the band diagrams under bias with Silvaco TCAD drift diffusion software at 0.0, −0.3, and −1.3 V. These band diagrams are plotted in Fig. 1 where the solid, dotted, and dashed curves correspond to 0.0, −0.3, and −1.3 V, respectively. The inset of Fig. 1 provides the CV measurement illustrated by the black dotted curve, while the best fit curve is plotted and shown by the gray curve. The CV fit range is shown by vertical lines and is chosen to be between −1.4 and 0.14 V. The fit range only extends to small forward biases (0.14 V) due to the nonuniform doping profile in the contact, while extending to larger reverse biases (−1.4 V) to best model the absorber layer doping. The best fit results to the measured CV provide inputs for plotting the band diagrams and are listed in Table I.[30] While the extracted doping concentration for the barrier layer is reported to be *n*-type in Table I, recent findings report that this may indicate an effective *n*-type concentration.[31] More specifically, the unintentionally doped Al-containing barrier is typically *p*-type[32,33] and may be compensated by interface donor-like defects introduced during growth. This detail is worth noting as it may have an impact on interpreting the extracted activation energies from DLTS that may be due to the barrier layer.

Based on the band diagrams illustrated in Fig. 1, an injection pulse of −0.3 V is near ideal as it minimizes the absorber depletion width while avoiding absorber accumulation, and the resulting band diagram resembles a typical majority carrier pulse in a $p^+n$ junction. Subsequently, a quiescent bias equal to −1.3 V is chosen to maximize the transient amplitude while limiting the magnitude of dark current across the largest possible temperature range to minimize the effects of capture via large dark currents on the DLTS spectra.[34,35] Figure 2 provides the voltage-dependent dark current density at temperatures ranging from 10 to 300 K and a corresponding capacitance transient at 14 K in the inset. Figure 2 demonstrates that tunneling currents in the InAs *nBn* begin to dominate the dark current from V = −1.1 to −1.5 V at temperatures <150 K. Using the inputs from Table I, an applied quiescent bias of −1.3 V results in an absorber potential drop of $\Phi_{AL} = -0.55$ V and, thus, an approximate depletion width of $W_D = \sqrt{2\epsilon_{AL}\epsilon_0\Phi_{AL}/qN_{D,AL}} = 309$ nm, where $\epsilon_{AL}$ is the absorber relative permittivity, $\epsilon_0$ is the free space permittivity, $q$ is the electron charge, and $N_{D,AL}$ is the absorber layer doping concentration.[30] While a larger depletion width would be more favorable for a quiescent bias condition, the device could be inundated from a large tunneling current magnitude; at a bias of −1.5 V the current density is roughly an order of magnitude larger than that at −1.3 V for temperatures <100 K. An injection pulse time of $\Delta t = 50$ ms and a quiescent bias dwelling time of 1 s is chosen for this study.

**TABLE I.** Relevant device input parameters used to simulate the 78 K band diagrams in Fig. 1 including values (in bold) determined from the best fit result of the CV measurement illustrated in the inset of Fig. 1. The contact layer doping is set to closely simulate true doping concentration (in parenthesis indicated by "grade") to best capture the true device doping profile in the band diagram. (f) next to the parameters denotes that the values are fixed in performing the CV fit to the data.

| | Absorber layer | Barrier layer | Contact layer |
|---|---|---|---|
| Materials | InAs | AlAsSb | InAs |
| $E_g$ (78 K) (eV) | 0.410 | 1.82 | 0.410 |
| Electrical type | *n* | *n* | *n* |
| $N_D$ ($cm^{-3}$) | **$9.72 \times 10^{15}$** | **$1.00 \times 10^{16}$** | **$1.03 \times 10^{16}$ (grade)** |
| Thickness ($\mu$m) | 3.85 | **0.104** | 0.150 |
| $N_c$ (300 K) ($cm^{-3}$) | $4.82 \times 10^{18}$ | | $4.82 \times 10^{18}$ |
| $N_v$ (300 K) ($cm^{-3}$) | $1.05 \times 10^{17}$ | | $1.05 \times 10^{17}$ |
| Electron affinity $\chi$ (eV) | 4.67 | 3.26 | 4.67 |
| Relative permittivity | **15.305 (f)** | **9.22** | **15.305 (f)** |

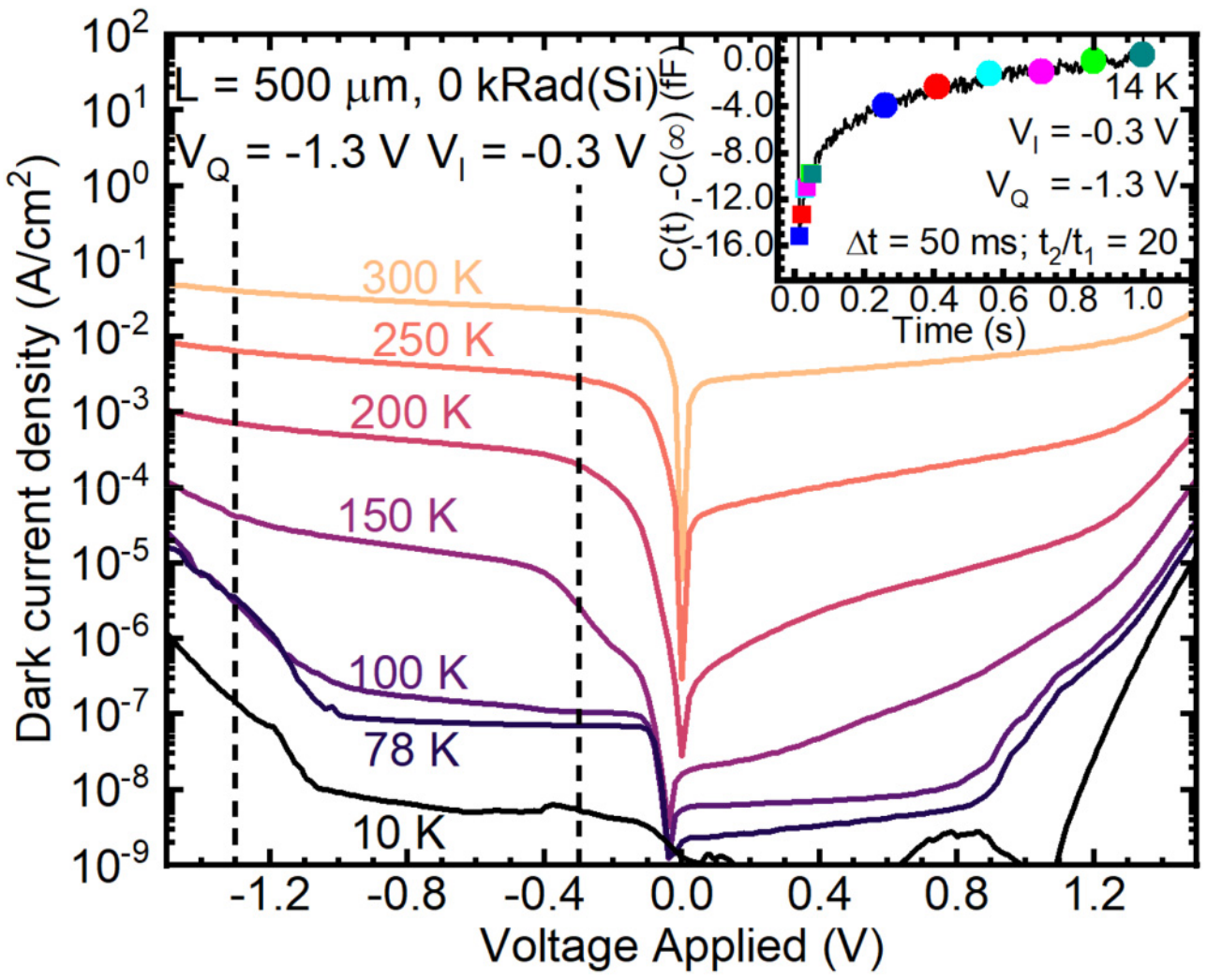

**FIG. 2.** Dark current densities of a 500 $\mu$m InAs *nBn* over the temperature range of interest (10–300 K), and inset, a capacitance transient at 14 K, all in pre-irradiation conditions. The injection ($V_I = -0.3$ V) and quiescent ($V_Q = -1.3$ V) biases are indicated by the vertical dashed lines. Inset transient results from an injection time of $\Delta t = 50$ ms and a quiescent voltage maintained ~1 s. The squares and circles in the inset transient represent the start and end of the color-coded rate windows used to generate the DLTS spectra. A rate window ratio of $t_2/t_1 = 20$ is used and $C(\infty) = C(t = 1$ s).

With the bias conditions established, the capacitance transient similar to that plotted in the inset of Fig. 2 can be measured, which is the average of 300 sequential transient measurements at 14 K. The change in capacitance, $\Delta C = C(t_1) - C(t_2)$, is evaluated to carry out a traditional rate window DLTS analysis. The $t_2$ and $t_1$ sample points are indicated by circles and squares in the capacitance transient (Fig. 2, inset), where each colored pair represents a unique rate window evaluation, and a ratio $t_2/t_1$ of 20 is utilized to maximize the amplitude $\Delta C$. For each rate window, the change in capacitance is plotted as a function of temperature (2 K steps), generating a spectrum. A defect peak or feature will occur in this spectrum and the temperature at which this feature occurs depends on the energy level of the defect and the rate window selected according to $\tau_e = e_n^{-1} = (t_1 - t_2)/\ln(t_1/t_2)$. Tracking the feature temperature location as a function of rate window yields the temperature and emission rate pairs that facilitate the Arrhenius analysis of the defect level. In this report, negative spectrum peaks and features correspond to majority carrier emission, while positive spectrum peaks and features correspond to minority carrier emission.

## B. Time-resolved photoluminescence

To relate the results extracted from DLTS to a secondary quantitative assessment of device optoelectronic quality, TRPL experiments are performed on an unprocessed (as-grown) piece of the InAs *nBn* to extract the minority carrier lifetime both as a function of temperature and high energy proton irradiation. The InAs sample is excited using a 1535 nm wavelength pulsed laser, and the photoluminescence signal is measured using a 6 $\mu$m cutoff VIGO systems PVI-4TE detector. The photoluminescence decays reported here are an average of 50 000 acquisitions from a Teledyne Lecroy HD 4096 oscilloscope. The TRPL is measured *in situ* between step doses of proton irradiation up to 100 kRad ($7.5 \times 10^{11}$ $H^+/cm^2$) and is plotted in the inset of Fig. 3. The minority carrier lifetime is extracted from low-excitation conditions (injection of $2 \times 10^{15}$ electron–hole pairs/cm$^3$ per pulse), where a single exponential decay rate is fit to the averaged signal.[11,12,17] The reciprocal of the measured minority carrier lifetime, the recombination rate constant $\tau^{-1}$, is extracted as a function of proton fluence and shown by the red squares in Fig. 3. The *in situ* lifetime damage factor $K_{1/\tau}$ is calculated as the slope to a least-squares linear fit to the data and is shown by the red line.

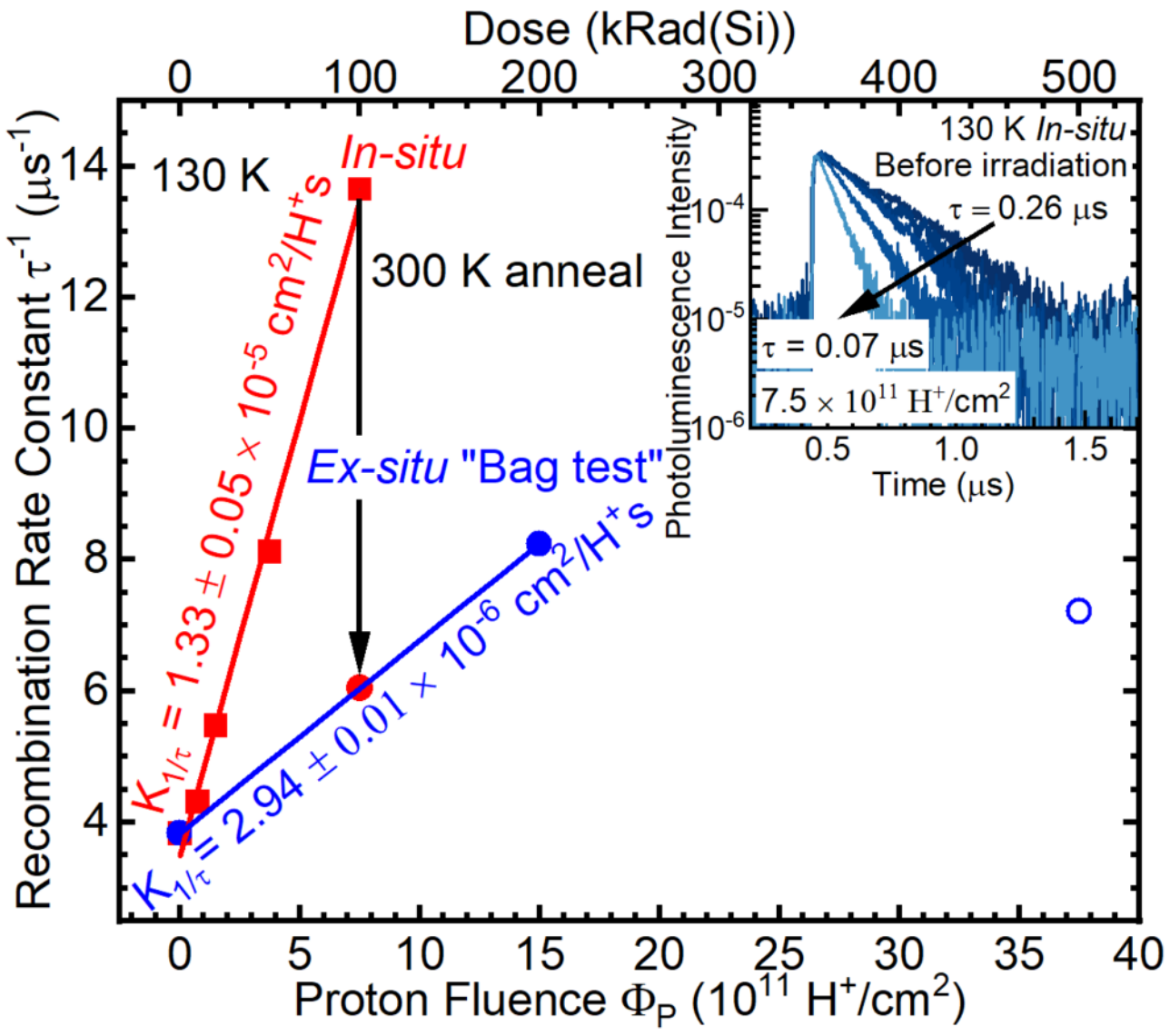

**FIG. 3.** *In situ* evaluation of the 130 K recombination rate constant (inverse of SRH lifetime) of the InAs *nBn* as a function of proton fluence and corresponding *in situ* TRPL decays (inset) from which the lifetimes are evaluated with dose. Red squares correspond to the *in situ* experiment, where the sample is maintained at 130 K during dosing, and a lifetime damage factor is determined from a linear fit (red line) to this data. The red circle corresponds to the recombination rate constant determined after a ~7 day room temperature anneal and the blue (both solid and hollow) data points correspond to the recombination rate constants extracted after room temperature dosing (so-called "bag testing"). The solid blue line corresponds to a linear fit to the red *and* blue solid circles, and the *ex situ* lifetime damage factor is provided. The inset provides the *in situ* TRPL decays in between step doses where the navy-blue curve is the unirradiated TRPL lifetime and lighter blue colors correspond to increasing proton irradiation.

The *in situ* irradiated InAs sample is then returned to room temperature and allowed to cool back to background radioactivity levels. The sample is then returned to 130 K, and TRPL is remeasured to extract the 130 K post-anneal recombination rate shown by the solid red circle in Fig. 3. Following the post-anneal measurement, the sample is cleaved into two separate pieces and sent back to UC Davis to attain two separate cumulative doses at 200 and 500 kRad(Si), and then their radioactivity is allowed to cool before further TRPL measurements are performed. The 200 and 500 kRad(Si) recombination rate constants are plotted by the solid and hollow blue circles, respectively. As the 100 kRad(Si) *in situ* post-anneal point (red circle) is effectively equivalent to a 100 kRad(Si) *ex situ* evaluation, a linear fit of only the solid circles [including the red circle representing the 100 kRad(Si) *in situ* post 300 K anneal recombination rate data point] is performed to determine a second lifetime damage factor that corresponds to the room temperature *ex situ* "bag test" experiment. The 500 kRad(Si) data point was excluded from the fit as it appears to be an outlier from the previous *ex situ* doses, possibly due to new species of defects being generated, convoluting the analysis, as discussed in Sec. III.

It is apparent here that room temperature dosing provides a significantly lower recombination rate damage factor (blue line) compared to low temperature dosing (red line) due to the defects generated having enough thermal energy at 300 K to be mobile enough to recombine, lowering the defect survival rate.[36,37] This annealing effect has been noted several times[9,38–42] and is a reminder that *ex situ* "bag test" evaluations of displacement damage should be treated with caution if the device under test is intended to operate in cryogenic environments.

## III. RESULTS AND DISCUSSION

### A. DLTS and emission study as a function of dose

Results of both the *ex situ* and *in situ* DLTS experiments can be found in Fig. 4, where the dose-dependent DLTS spectra at the 82 ms emission time constant and the resulting Arrhenius analysis are shown. The DLTS spectra in Figs. 4(a) and 4(b) are analyzed by simulating a minimal number of Gaussians to best capture the line shape of the spectra [e.g., at 0 kRad(Si), a sum of two Gaussians and a uniform background were used to best simulate the spectrum] and best-fit Gaussian centers are then taken as inputs for determining the temperature–emission rate pairs for the Arrhenius analysis in Figs. 4(c) and 4(d). It can be seen at the 0 kRad(Si) condition [dark blue curves in (a) and (b)] that there are two distinct features: one that appears to be a broad shoulder below 75 K and then a high temperature minimum at ~240 K. Because the low temperature shoulder at 0 kRad(Si) does not resemble a true Gaussian peak, the low temperature fit windows for the different rate windows were tuned to best capture the shoulder, providing an upper limit to the activation energy of $E_1^{0\ kRad(Si)} = 29$ meV. Presumably, given the bias and doping conditions of the absorber, this is a shallow electron defect and relative to the InAs conduction band. Speculating, this shoulder feature's broadness suggests that there may be a spectrum of defects manifesting here due to the Si dopant complexing with a native shallow intrinsic defect. A variety of predictions from tight-binding and density functional theory calculations have labeled potential shallow defects in InAs to be an In vacancy $V_{In}$,[43] an As vacancy $V_{As}$,[44] or an In interstitial $In_i$.[45] Positron annihilation lifetime spectroscopy studies on InAs substrates have found evidence for vacancy complexes in both as-grown substrates and as a function of proton irradiation.[46,47]

For the local minimum that appears at ~240 K, 82 ms time constant, a pre-irradiation activation energy of $E_{deep}^{0\ kRad(Si)} = 539$ meV is extracted. This feature presents as either a defect above the InAs conduction band or a mid-level gap defect in the wider bandgap AlAsSb barrier material. This high activation energy feature only occurs at temperatures where the CV measurements indicate that the barrier is no longer fully depleted at low reverse biases,[30,48] suggesting that this deep level is more likely a defect in the barrier material itself. Interestingly, the AlAs system is predicted to have an Al vacancy $V_{Al}$ at 540 meV relative to the valence band.[43] Thus, in pre-irradiation conditions, two independent features in the DLTS spectra with very different natures are observed. Determining whether these defect amplitudes grow as a function of 63 MeV proton irradiation, or if new spectrum features arise with dose will provide evidence as to whether proton irradiation increases these intrinsic defect concentrations or incorporates new species of defects.

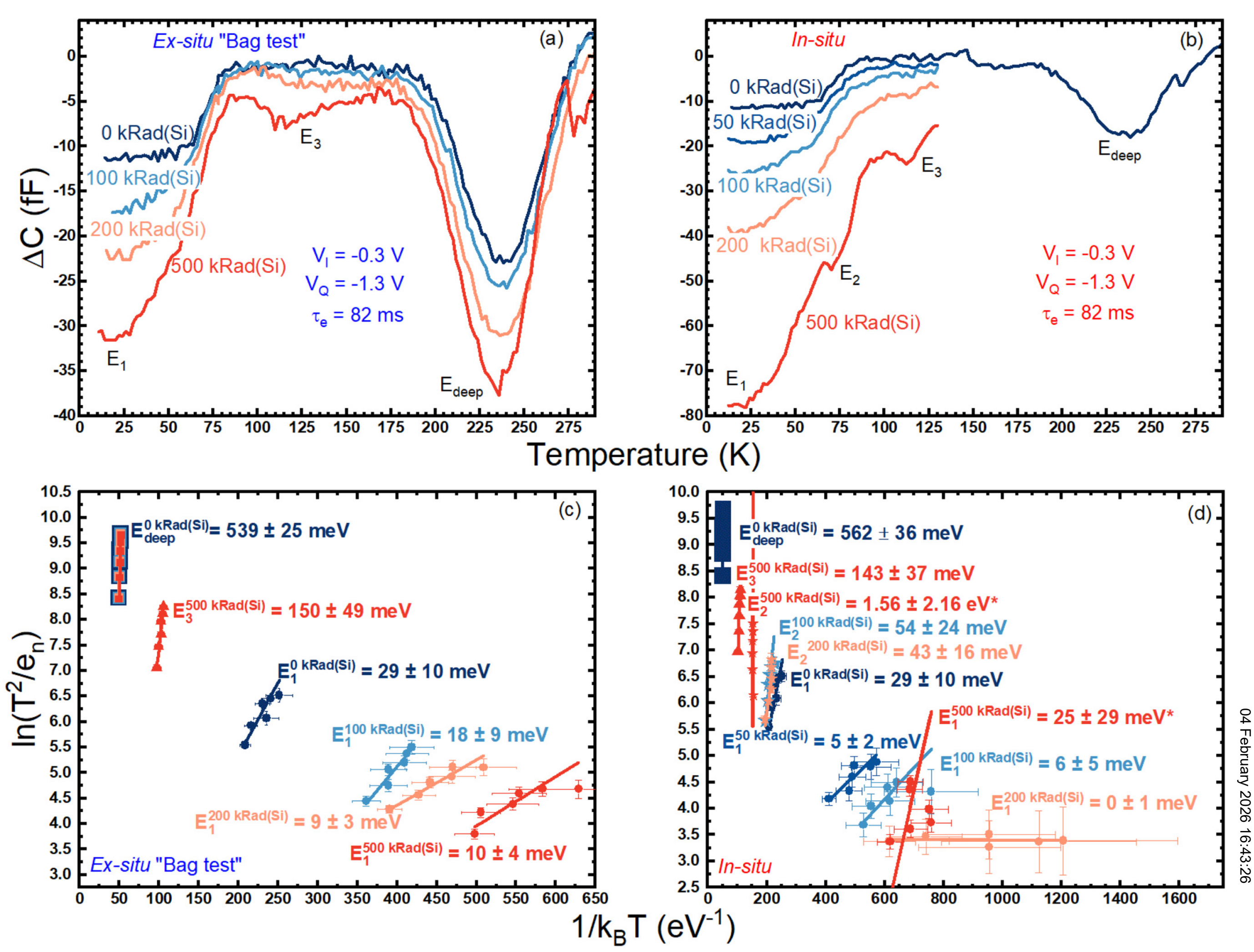

**FIG. 4.** Dose-dependent DLTS spectra (82 ms emission time constant) and a corresponding Arrhenius analysis of InAs *nBn*s for the *ex situ* [(a) and (c)] and *in situ* [(b) and (d)] experiments. For (b), following each step dose at 10 K, transients were acquired as a function of temperature up to 130 K to generate the spectra indicated, and then cooled back down to 10 K for the subsequent dose. The asterisks in subplot (d) indicate a lower degree of confidence in the extracted defect levels due to the complexity of the 500 kRad(Si) spectrum illustrated in subplot (b).

Upon examining the spectra as a function of dose in Fig. 4, the amplitudes for the shallow and deep levels observed in the pre-irradiation conditions appear to grow monotonically as a function of both room temperature [*ex situ*, (a)] and 10 K [*in situ*, (b)] 63 MeV proton irradiation. For both cases, it appears that the shallow shoulder becomes more peak-like with increasing dose, allowing for the fit range to extend to all of the cryogenic temperatures after the first 63 MeV proton irradiation dose. This results in a consistent trend toward smaller activation energies for the *ex situ* experiment [see Fig. 4(a) and 4(c)], suggesting that room temperature dosing imparts a shallow defect with a single energy level that begins to dominate with increasing dose. However, for the *in situ* experiment [see Fig. 4(b) and 4(d)], it appears that the defect trends down to 0 meV more quickly, after only the 200 kRad(Si) dose, and then after the 500 kRad(Si) dose, the activation energy uncertainty becomes larger than the value itself, suggesting that this simple analysis is no longer sufficient. It can also be seen after 100 kRad(Si) that another defect $E_2$ begins to appear between 50 and 75 K with an activation energy of 54 meV. This becomes a stronger feature after 500 kRad(Si), followed by a new feature $E_3$ after the 500 kRad(Si) dose. Although the emission from these three defect features is occurring at three discernable nearby temperatures, their proximity to each other may be affecting the validity of the analysis, warranting the need for an inverse Laplace transform on the transients to improve fidelity.[49] An in-depth Laplace transform study on these transients will be the subject of a future report to further understand the emission spectrum signature of these transients.

With the transients extracted, it would be instructive to examine both the doping concentration and the transient amplitude as a function of both *in situ* and *ex situ* proton irradiation to determine the shallow defect introduction rate. A simple Schottky approximation CV analysis from −0.5 to −1.0 V is performed to provide a valid assessment of the trends of the *nBn*s' doping concentration as a function of dose.[30] Coupled with the extracted doping concentrations, the *nBn*s' dose-dependent trap concentration $N_T$ is calculated via[18]

$$N_T = 2\frac{\Delta C_\infty}{C}(N_D - N_A). \tag{1}$$

In Eq. (1), $\Delta C_\infty$ is the capacitance change due to the injected majority pulse ($C(\infty) - C(0)$), $C$ is the capacitance of the *nBn* under the quiescent reverse bias condition ($C(\infty)$), and $N_D - N_A$ is the net doping concentration extracted from the CV analysis. The *nBns*' doping and corresponding defect introduction rate are plotted in Figs. 5(a) and 5(b), respectively. It can be seen in Fig. 5(a) that the InAs *nBn*s have an initial doping of $9.77 \times 10^{15}\ \text{cm}^{-3}$ and that there is a noticeable linear increase in doping concentration for *in situ* (red squares, and trendline) high energy proton irradiation with an extracted doping addition rate of $105 \pm 12\ \text{cm}^{-1}/\text{H}^+$. This rate of increase appears to be nearly identical to the results from a previous *in situ* CV study on a mid-wave infrared *nBn* with an undoped absorber concentration of $\sim 1.5 \times 10^{14}\ \text{cm}^{-3}$,[39] but may be difficult to resolve in cases where the material is intentionally doped to $\sim 4 \times 10^{15}\ \text{cm}^{-3}$ and dosed up to ≤100 kRad(Si) ($\leq 7.5 \times 10^{11}\ \text{H}^+/\text{cm}^2$).[9,42] In contrast to the *ex situ* study, no clear trends are apparent and may be unresolvable as shown in blue, a phenomenon similarly reported in Ref. 50.

With the doping concentrations extracted from the CV measurements, Eq. (1) enables evaluation of trap concentration and thus, a trap introduction rate can now be extracted. Here, a single trap species is assumed in the low temperature regime for the spectra seen in Figs. 4(a) and 4(b), where the corresponding transients (which yield the factor $\Delta C_\infty/C$) as a function of *in situ* dose can be seen in the inset of Fig. 5(b). The main plot in Fig. 5(b) provides the extracted 18 K trap concentrations as a function of dose for the InAs device dosed *in situ* (red squares) and the device dosed *ex situ* (blue circles). A linear fit is then performed on the data to determine a trap introduction rate $dN_T/d\Phi_P$ for each experiment. The results indicate the *in situ* trap introduction rate is >4× the *ex-situ* trap introduction rate. It is found after the 500 kRad DLTS sweep, the 18 K transient amplitude changed by <10%, indicating that this *in situ* trap introduction rate is partially attributable to DLTS measurement-induced annealing. This significant difference highlights the importance of performing a material test such as this *in situ,* given the expectation that ultimately the focal plane arrays fabricated from these materials will always be held at low temperatures during operation in space. Furthermore, it is enlightening to see just how much thermal annealing occurs in the *ex situ* irradiation experiment compared to the *in situ* irradiation experiment. Based on these defect introduction rates, it is calculated that roughly 77% of the proton irradiation induced defects are annealed due to room temperature annealing in the InAs material.

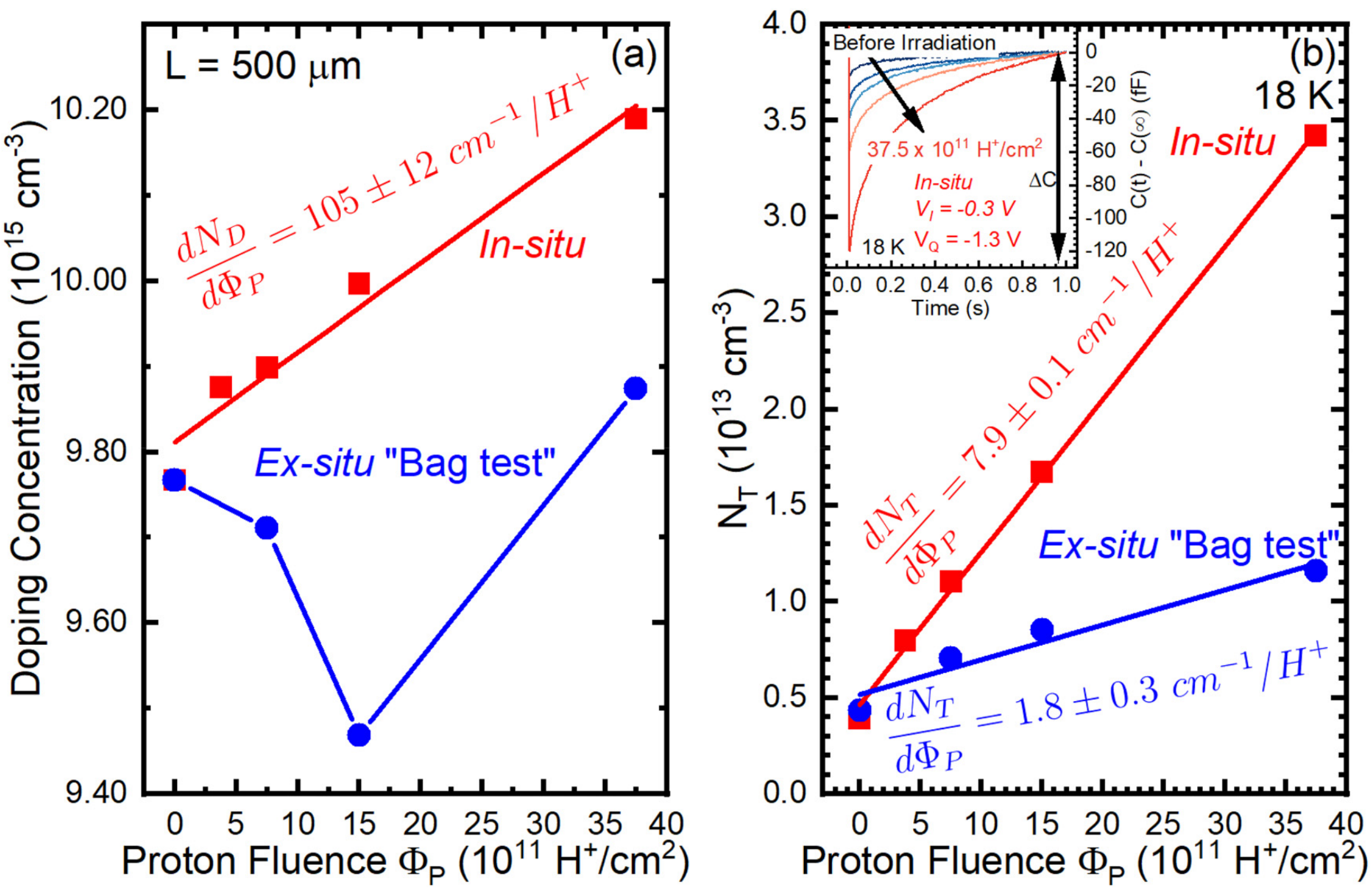

**FIG. 5.** Extracted doping (a) and shallow defect (b) concentrations as a function of 63 MeV proton fluence for the InAs *nBn* from the *in situ* (red data) and *ex situ* (blue data) experiments, using 10 K and room temperature irradiation, respectively. Calculated 18 K trap concentrations and trap introduction rates as a function of *in-situ* (red squares and line) and *ex situ* proton irradiation (blue squares and line). The inset to (b) plots the 18 K transients that correspond to the *in situ* assessment of the trap concentration from 0 kRad(Si) to 500 kRad(Si) ($37.5 \times 10^{11}\ \text{H}^+/\text{cm}^2$).

## B. Recombination rate analysis as a function of dose

The temperature-dependent minority carrier lifetime $\tau_{mc}$ evaluated from the low-injection photoluminescence decay is analyzed as a sum of the recombination rates of the individual lifetime mechanisms,

$$\frac{1}{\tau_{mc}} = \frac{1}{\tau_{SRH}} + \frac{1}{\phi \tau_{rad}} + \frac{1}{\tau_{Auger}}. \tag{2}$$

In Eq. (2), $\tau_{SRH}$ is the SRH lifetime,[1] $\tau_{rad}$ is the radiative lifetime scaled by the photon recycling factor $\phi$,[15,51,52] and $\tau_{Auger}$ is the Auger lifetime.[13,14] The SRH recombination process describes the recombination of photogenerated carriers that recombine at defect centers (either lattice impurities or crystalline defects) with energy levels located within the bandgap of the material. It is a function of the electron and hole lifetimes $\tau_{n0}$, and $\tau_{p0}$ that are then scaled by the material's electron and hole concentrations $n_0$, and $p_0$, and characteristic concentrations, $n_1$ and $p_1$, expressed by

$$\tau_{SRH} = \frac{\tau_{p0}(n_0 + n_1) + \tau_{n0}(p_0 + p_1)}{n_0 + p_0}. \tag{3}$$

In Eq. (3), the SRH recombination can be increased (minority carrier lifetime decreased) if more defects are present in the material, and this relation of defect concentration is found by examining the electron and hole lifetimes,

$$\frac{1}{\tau_{p0}} = \sigma_p v_p N_T, \quad \frac{1}{\tau_{n0}} = \sigma_n v_n N_T. \tag{4}$$

It can be seen through Eq. (4) that the electron and hole lifetimes are dependent on the trap concentration $N_T$, recombination cross section $\sigma_n$ and $\sigma_p$, and thermal velocities $v_n$ and $v_p$. Since the lifetimes depend on the product of the trap concentration and cross section, the two quantities cannot be extracted in isolation without assumptions. The SRH's dependence on the trap defect levels can be found by examining the characteristic concentrations $n_1$ and $p_1$ in Eq. (5),

$$n_1 = N_c \exp\left(\frac{-(E_c - E_T)}{k_B T}\right), \quad p_1 = N_v \exp\left(\frac{-(E_T - E_v)}{k_B T}\right). \tag{5}$$

The characteristic concentration $n_1$ ($p_1$) is the density of electrons (holes) in the conduction (valence) band with effective density of states $N_c$ ($N_v$) for the case that the Fermi level is located at the defect energy level $E_T$. When analyzing the temperature-dependent minority carrier lifetime, the material's doping concentration ($n_0 - p_0$), trap level relative to the conduction band ($E_c - E_T$), and trap concentration cross section product ($\sigma_p N_T = \sigma_n N_T = \sigma N_T$) are set as fit parameters to the data.

The materials' radiative recombination $\tau_{rad}$ is calculated by

$$\tau_{rad} = \frac{n_i^2}{G_r(n_0 + p_0)}, \tag{6}$$

where $n_i$ is the intrinsic carrier concentration and $G_r$ is the integrated thermal emission spectrum, where its definition and dependence can be found in Refs. 12 and 15. Then, with this $n$-type InAs, the Auger-1 process $\tau_{A1}$ is assumed, where it's defined by[12–14]

$$\tau_{Auger} = \frac{2n_i^2}{n_0^2 + n_0 p_0} \times \tau_{A1}. \tag{7}$$

It is found with this material's $\sim 1 \times 10^{16}\ \mathrm{cm^{-3}}$ doping concentration and larger bandgap in comparison with 5 $\mu$m (∼250 meV) material, the recombination rate analysis is insensitive to the parameters of the Auger lifetime as can be seen in Fig. 6, where the Auger component is greater than an order of magnitude above the determined lifetimes and off the scale of the plot. The increasing lifetime with increasing temperature is a property resultant from the shallow SRH energy level in this material.

Figure 6 provides the temperature-dependent minority carrier lifetime as a function of *ex situ* dose from 0 to 37.5 $\mathrm{H^+/cm^2}$ [500 kRad(Si)], where a recombination rate analysis is performed for the two endpoints of dose and is shown by the solid navy-blue (before irradiation) and red [500 kRad(Si)] curves. The SRH analysis indicates that $E_T = 10$ meV and $E_T = 1$ meV for these two endpoints, respectively, consistent with the observation of an increasingly more shallow defect level with increasing dose in DLTS. Both the SRH and radiative lifetime components that make up the pre-irradiation lifetime fit are shown by the dotted–dashed and dashed curves, respectively. The doping concentrations in this fit are fixed at the extracted doping concentrations determined from the CV measurements. It can be seen for all doses that the minority carrier lifetime increases with temperature up to room temperature, a behavior that is not typical in previous works on mid-wave infrared materials, where extracted defect levels were closer to mid-bandgap.[4,12,16,53,54] The extracted effective defect energies seen in this work appear to be shallow toward the conduction band in all cases. It is also worth noting that a single defect appears to be sufficient to achieve a good fit, even at the highest *ex situ* dose where the $E_3$ defect appears at ∼130 K in Fig. 4(a).

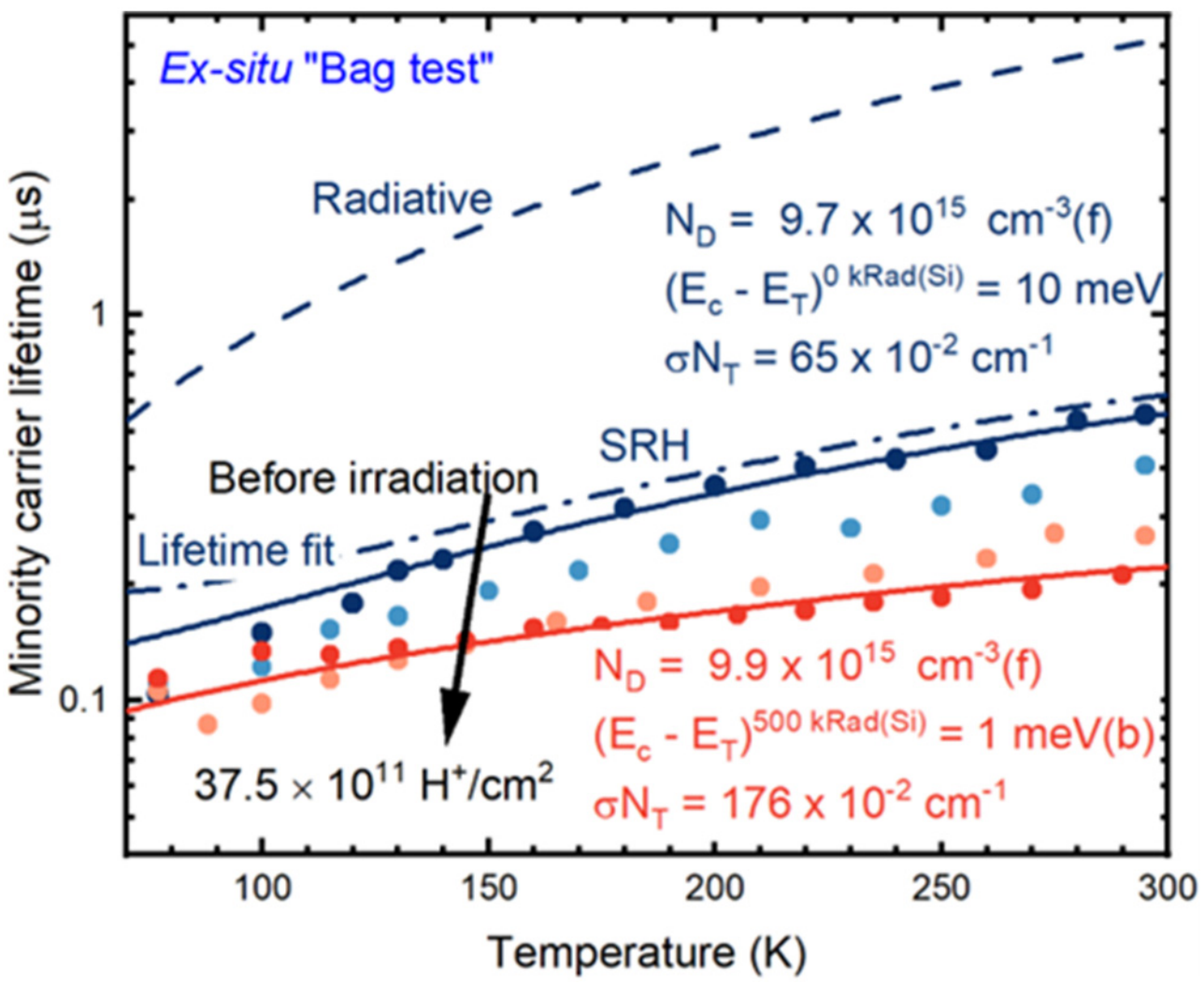

**FIG. 6.** Low injection temperature-dependent minority carrier lifetime of the InAs *nBn* as a function of *ex situ* dosing. The color gradient is from 0 $H^+/cm^2$ (navy blue) to $37.5 \times 10^{11}$ $H^+/cm^2$ [500 kRad(Si), red] proton irradiation, where a temperature-dependent recombination rate analysis is performed on the pre-irradiation and post-irradiation conditions. The black arrow is a guide to show the decrease in lifetime with increasing irradiation. The SRH and the radiative lifetime components to the pre-irradiation data (navy-blue circles) are shown by the navy-blue dotted–dashed and dashed curves, respectively. The lifetime fit, which is the sum of the recombination rates, is shown by solid navy-blue curve and solid red curve for the pre-irradiation and post-irradiation [$37.5 \times 10^{11}$ $H^+/cm^2$, 500 kRad(Si)] conditions. The corresponding fit results for the two end point irradiation conditions are provided. The "(f) " next to the parameters indicates that the inputs are fixed to the CV determined doping concentration and the "(b) " next to the individual parameter indicates that the fit is at the parameter bounds.

This confirms that a recombination rate analysis can extract an effective defect energy that can fit to the lifetime results but does not sufficiently capture the characteristics of any multitude of defects that are prevalent in the material. DLTS, or some complementary evaluation, is therefore necessary to assess the material's true defect nature.[16]

Nevertheless, given the extracted defect for the pre-irradiation condition is shallow and provides some agreement to the DLTS result, a defect capture cross section can be determined. This assumes the shallow trap concentration extracted from the 18 K emission transient in Fig. 5(b) to be equivalent to the recombination defect concentration. Provided $\sigma N_T = 65 \times 10^{-2}$ $\text{cm}^{-1}$ and $N_T^{0\ kRad(Si)} = 3.9 \times 10^{12}$ $\text{cm}^{-3}$, the calculated cross section is $\sigma^{0\ kRad} = 1.6 \times 10^{-13}$ $\text{cm}^2$, assuming both the electron and hole capture cross sections are the same ($\sigma_p = \sigma_n$). It is worth noting that this analysis can only be justified when DLTS provides a result that agrees with the recombination rate analysis (single defect from both measurements, providing similar defect level/activation energy values). Further studies are required to separate the electron and hole capture cross sections, such as an injection pulse width study to determine majority capture times.[55]

## IV. CONCLUSION

In summary, an InAs *nBn* detector is characterized by DLTS and TRPL to examine both the generation and recombination dynamics in the material. Optimal injection and quiescent bias conditions are selected by examining both the CV and dark current of the device. It is found that a possible spectrum of shallow defects may be present in the *nBn* evidenced by the low temperature broad shoulder seen in the pre-irradiation DLTS spectra, and that a defect with a wider-than-bandgap activation energy manifests at higher temperatures. It is proposed that this deep defect may be emission from a defect in the barrier layer itself. The experiments are performed as a function of both *in situ* and *ex situ* 63 MeV proton irradiation to ascertain the level of thermal annealing on the defect generation as well as determine whether proton irradiation increases the concentration of the intrinsic, pre-irradiation defects, or whether new defect species are incorporated. It is found in the low <200 K range, proton irradiation introduces three noticeable features in the DLTS spectra, where one feature appears to be intrinsic to this Si-doped InAs *nBn* at low temperatures (the low temperature shoulder). As the proton irradiation increases to higher cumulative doses, the results provided by the rate window method of DLTS appear to become suspect due to the broadness and proximity of the low temperature features. The device dosed *in situ* shows a steady increase in doping concentration with increasing dose, but this change becomes obscured by room temperature annealing, and the simple systematic uncertainty for the case of the device dosed *ex situ*. Following the doping characterization, a trap introduction rate is extracted, and a ∼4× factor between the room temperature *ex situ* and cryogenic *in situ* defect introduction rate is observed, justifying the need for a correspondence between *ex situ* and *in situ* studies for accurate predictions of device performance on space-based mission settings.

## ACKNOWLEDGMENTS

The authors thank Eduardo Garcia for device fabrication support, Perry C. Grant for experimental assistance, and Gyorgy Vizkelethy, John M. Cain, and Devika Mehta for helpful conversations. The authors acknowledge financial support through research sponsored by the Air Force Research Laboratory, Section 219 Seedling for Disruptive Capabilities, and the Air Force Office of Scientific Research (Project No. 22RVCOR016). This work was performed, in part, at the Center for Integrated Nanotechnologies, an Office of Science User Facility operated for the U.S. Department of Energy (DOE) Office of Science. This work was partially supported by the Laboratory Directed Research and Development Program at Sandia National Laboratories under project 233119. This article has been authored by an employee of National Technology & Engineering Solutions of Sandia, LLC, under Contract No. DE-NA0003525 with the U.S. Department of Energy (DOE). The employee owns all right, title, and interest in and to the article and is solely responsible for its contents. The United States Government retains and the publisher, by accepting the article for publication, acknowledges that the United States Government retains a non-exclusive, paid-up, irrevocable, world-wide license to publish or reproduce the published form of this article or allow others to do so, for United States Government purposes. The DOE will provide public access to these results of federally sponsored research in accordance with the DOE Public Access Plan at https://www.energy.gov/downloads/doe-public-access-plan. The views expressed in the article do not necessarily represent the views of the U.S. Department of Energy, or the official policy of position of the Department of the Air Force, the Department of Defense, or the United States Government. Approved for public release: distribution is unlimited. Public affairs release approval number: 2025-5369.

## AUTHOR DECLARATIONS

### Conflict of Interest

The authors have no conflicts to disclose.

### Author Contributions

**Rigo A. Carrasco:** Conceptualization (lead); Formal analysis (lead); Investigation (lead); Software (equal); Writing – original draft (lead); Writing – review & editing (equal). **Christopher P. Hains:** Data curation (equal); Investigation (equal); Writing – review & editing (equal). **Nathan Gajowski:** Data curation (equal); Formal analysis (equal); Investigation (equal); Writing – review & editing (equal). **Alexander T. Newell:** Data curation (equal); Software (equal); Visualization (equal); Writing – review & editing (equal). **Julie V. Logan:** Data curation (equal); Investigation (equal); Writing – review & editing (equal). **Zinah M. Alsaad:** Investigation (equal); Software (supporting); Writing – review & editing (supporting). **Preston T. Webster:** Conceptualization (equal); Investigation (equal); Resources (equal); Supervision (equal); Visualization (equal); Writing – original draft (supporting); Writing – review & editing (equal). **Christian P. Morath:** Conceptualization (equal); Formal analysis (supporting); Project administration (equal); Resources (equal); Supervision (equal); Writing – review & editing (equal). **Diana Maestas:** Funding acquisition (equal); Project administration (equal); Resources (equal); Supervision (equal); Writing – review & editing (equal). **Aaron J. Muhowski:** Methodology (equal); Resources (equal); Writing – review & editing (equal). **Samuel D. Hawkins:** Data curation (equal); Methodology (equal); Resources (equal); Writing – review & editing (equal). **Evan M. Anderson:** Conceptualization (equal); Investigation (equal); Project administration (equal); Resources (equal); Supervision (equal); Visualization (equal); Writing – review & editing (equal).

## DATA AVAILABILITY

The data that support the findings of this study are available from the corresponding author upon reasonable request.